\documentclass[12pt]{spieman}  
\usepackage{amsmath,amsfonts,amssymb}
\usepackage{graphicx}
\usepackage{setspace}
\usepackage{tocloft}
\usepackage{subcaption}
\usepackage{booktabs}
\usepackage{lineno}
\usepackage{gensymb}
\usepackage{bm}
\usepackage{xcolor}
\usepackage{placeins}

\title{Impact of deep learning-based image super-resolution on binary signal detection}

\author[a,$\dagger$]{Xiaohui Zhang}
\author[b,$\dagger$]{Varun A. Kelkar}
\author[c]{Jason Granstedt}
\author[a,d,e]{Hua Li}
\author[a,b,c,*]{Mark A. Anastasio}
\affil[a]{University of Illinois at Urbana-Champaign, Department of Bioengineering, Urbana, IL 61801, USA}
\affil[b]{University of Illinois at Urbana-Champaign, Department of Electrical and Computer Engineering, Urbana, IL 61801, USA}
\affil[c]{University of Illinois at Urbana-Champaign, Department of Computer Science,  Urbana, IL 61801, USA}
\affil[d]{University of Illinois at Urbana-Champaign, Cancer Center at Illinois, Urbana, IL 61801, USA}
\affil[e]{Carle Cancer Center, Carle Foundation Hospital, Urbana, IL 61801, USA}
\affil[$\dagger$]{Equal contribution}

\renewcommand{\vec}[1]{\mathbf{#1}}

\newcommand{\clearsubcaptcounter}{\setcounter{sub\@captype}{0}}

\renewcommand{\vec}[1]{\mathbf{#1}}

\newcommand{\f}{\mathbf{f}}
\newcommand{\g}{\mathbf{g}}

\renewcommand{\vec}[1]{\ensuremath{\mathbf{#1}}}

\def\[{\begin{equation}}
\def\]{\end{equation}}
\DeclareMathOperator*{\argmin}{arg\,min}

\DeclareMathAlphabet\mathbfcal{OMS}{cmsy}{b}{n}

\def\equationautorefname~#1\null{Eq.~(#1)\null}

\cftpagenumbersoff{figure}
\cftpagenumbersoff{table} 
\begin{document} 
\maketitle

\begin{abstract}

\noindent\textbf{Purpose:} Deep learning-based image super-resolution (DL-SR) has shown great promise in medical imaging applications. To date, most of the proposed methods for DL-SR have only been assessed by use of traditional measures of image quality (IQ) that are commonly employed in the field of computer vision. However, the impact of these methods on objective measures of image quality that are relevant to medical imaging tasks remains largely unexplored. In this study, we investigate the impact of DL-SR methods on binary signal detection performance.

\noindent\textbf{Approach:} Two popular DL-SR methods, the super-resolution convolutional neural network (SRCNN) and the super-resolution generative adversarial network (SRGAN), were trained by use of simulated medical image data. Binary signal-known-exactly with background-known-statistically (SKE/BKS) and signal-known-statistically with background-known-statistically (SKS/BKS) detection tasks were formulated. Numerical observers, which included a neural network-approximated ideal observer and common linear numerical observers, were employed to assess the impact of DL-SR on task performance. The impact of the complexity of the DL-SR network architectures on task-performance was quantified. In addition, the utility of DL-SR for improving the task-performance of sub-optimal observers was investigated. 

\noindent\textbf{Results:} Our numerical experiments confirmed that, as expected, DL-SR could improve traditional measures of IQ. However, for many of the study designs considered, the DL-SR methods provided little or no improvement in task performance and could even degrade it. It was observed that DL-SR could improve the task-performance of sub-optimal observers under certain conditions.

\noindent\textbf{Conclusions:} The presented study highlights the urgent need for the objective assessment of DL-SR methods and suggests avenues for improving their efficacy in medical imaging applications.
\end{abstract}


{\noindent \footnotesize\textbf{*}Mark A. Anastasio,  \linkable{maa@illinois.edu} }

\begin{spacing}{1}   

\section{Introduction}
\label{sect:intro}  
Single image super-resolution (SISR) is a classic image restoration operation that seeks to estimate a high-resolution (HR) image from an observed low-resolution (LR) one\cite{sisr-review}. 
A variety of methods have been developed to achieve this goal, 
such as filtering and interpolation-based approaches \cite{interpolation} and more formal regularized inverse problem-based formulations \cite{edge_prior, candes}, to name a few.
Recently, deep learning-based image super-resolution (DL-SR) methods have been widely employed and have shown great promise for SISR in terms of traditional image quality (IQ) metrics such as mean square error (MSE), structural similarity index metric (SSIM) and peak signal-to-noise ratio (PSNR)\cite{dl-sisr-review, srcnn,lai2018fast,srgan}.
%

In medical imaging, images are often acquired for specific purposes and the use of objective measures of IQ is widely advocated for assessing imaging systems and image processing algorithms\cite{barrett,mi-model-obs, snr-mi-system, mi-assess-iq, consensus-mi-system,weimin_cnnio,kaiyan-denoise}. 
Although DL-SR algorithms can improve traditional IQ metrics,
\cite{mammogram-sr, gancircle, gan-circle-mri-ct, deepresolve,ma2020mri, sr-hrme} 
it is well-known that such metrics may not always correlate with objective task-based IQ measures.
\cite{ct-recon-iq-index, model-obs-iq, noise-detectability-myers,vct-task-based} 
Despite this, relatively few studies have objectively assessed image super-resolution methods \cite{wang2020deep, sr-visiontask,sr-task-eval, deepresolve}. Dai \textit{et al.} evaluated six image super-resolution methods on popular vision tasks such as edge detection and semantic image segmentation and found that the standard perceptual metrics correlated well with the usefulness of image super-resolution to these tasks  \cite{sr-visiontask}.
Jaffe \textit{et al.} conducted a study in which the aesthetic image quality that DL-SR methods sought to improve did not necessarily increase classification accuracy \cite{sr-task-eval}. However, none of these studies were carried out with images, tasks or observers relevant to medical imaging. Additionally, the data processing inequality indicates that the performance of an ideal observer on a particular task cannot be improved by use of image processing transformations \cite{dpi}. The scenarios under which DL-SR may improve the performance of a sub-optimal observer on a specified task have not been thoroughly investigated. 

The purpose of this work is to evaluate DL-SR methods by use of task-based measures, as a preliminary attempt to address the issues raised above.
For this study, two canonical DL-SR networks were identified for the analysis. A variety of mathematical and learning-based numerical observers (NOs) were computed on the HR images, the LR images, as well as the images resolved by the DL-SR methods. Receiver operating characteristics (ROC) analysis was employed to quantify the performance of these NOs. Two {stylized} binary signal detection tasks were designed to evaluate the DL-SR networks systematically and comprehensively under known statistical conditions. Specifically, a {signal-known-exactly and background-known-statistically} (SKE/BKS) Rayleigh discrimination task\cite{rayleigh_ct, rayleigh_recon} was employed to assess the ability of a DL-SR to resolve two small adjacent objects. The inherent detectability of the signal was varied and its effect on the utility of DL-SR for improving detection task performance was studied. The impact of the depth of a DL-SR network on NO performance was investigated to see if the deep learning mantra "deeper is better" holds true for signal detection performance \cite{deeper-better}. Additionally, a {signal-known-statistically and background-known-statistically} (SKS/BKS) microcalcification (MC) cluster detection task was employed to investigate under what circumstances DL-SR techniques may improve the binary signal detection performance of a sub-optimal observer. 

The remainder of this paper is organized as follows. \autoref{sec:bkd} describes the relevant background on linear imaging systems, the basic theory relating to binary signal detection tasks, numerical observers and deep learning-based image super-resolution. \autoref{sec:numerical} describes the setup for the numerical studies and \autoref{sec:results} describes the results of the proposed evaluation. \autoref{sec:discussion} presents a discussion on the salient findings and \autoref{sec:conclusion} concludes the paper.

\section{Background}\label{sec:bkd}
Many imaging systems can be approximately described by a continuous-to-discrete (C-D) linear imaging model \cite{barrett}:
\begin{linenomath}
\begin{align}
    \g = \mathcal{H}f(\vec{r}) + \vec{n},
\end{align}
\end{linenomath}
where $f\in\mathbb{L}_2(\mathbb{R}^d)$ is the true object of interest that is a function of the $d$-dimensional spatio-temporal coordinate $\vec{r}$, and $\g \in \mathbb{E}^m$ is a vector that describes the measurement data. The mapping $\mathcal{H} : \mathbb{L}_2(\mathbb{R}^d) \rightarrow \mathbb{E}^m$ denotes the continuous-to-discrete (C-D) forward operator that represents the data-acquisition process, and $\mathbf{n} \in \mathbb{E}^m$ denotes the measurement noise. In practice, discrete-to-discrete (D-D) models for the imaging system are often employed, in which case the object $f(\vec{r})$ is approximated by a vector $\f \in \mathbb{E}^n, ~ n\in \mathbb{N}$, and a D-D approximation $\vec{H} \in \mathbb{E}^{m\times n}$ is employed in place of $\mathcal{H}$ \cite{barrett}.

\par

\subsection{Binary signal detection tasks}
A binary signal detection task requires an observer to classify the image as satisfying either hypothesis $H_0$ or hypothesis $H_1$:
\begin{linenomath}
\begin{align}
    H_0 &: \g = \vec{H}\f_0 + \vec{n} = \vec{H}(\f_b + \f_{s_0}) + \vec{n},\\
    H_1 &: \g = \vec{H}\f_1 + \vec{n} = \vec{H}(\f_b + \f_{s_1}) + \vec{n},
\end{align}
\end{linenomath}
where $\f_b \in \mathbb{E}^n$ denotes the background, $\f_{s_0}\in \mathbb{E}^n$ and $\f_{s_1}\in \mathbb{E}^n$ represent the signal under the two hypotheses, $\vec{H} \in \mathbb{E}^{m\times n}$ refers to the D-D imaging operator and $\mathbf{n}\in \mathbb{E}^m$ denotes the measurement noise. The special case where $\f_{s_0} = \vec{0}$ corresponds to a task of detecting the presence or absence of the signal $\mathbf{f}_{s_1}$ in an image. When $\f_b$ is a random vector drawn from a certain non-degenerate distribution and $\f_{s_0}$ and $\f_{s_1}$ are fixed known signals, the detection task is known as a {signal-known-exactly and background-known-statistically} (SKE/BKS) detection task. Alternatively, if $\f_{s_0}$ and $\f_{s_1}$ are also random, then the detection task is known as a {signal-known-statistically and background-known-statistically} (SKS/BKS) detection task. Both these tasks are considered in this work. 

\subsection{Numerical observers for IQ assessment}\label{NOs review}
A numerical observer (NO) for a signal detection task maps a given set of measurements $\g$, or alternatively, an image estimate $\hat{\f} \in \mathbb{E}^n$ of the object obtained from $\g$, to a scalar test statistic $t$ that is used to determine whether $\g$ or $\hat{\f}$ satisfies $H_0$ or $H_1$ based on comparison with a predetermined threshold $\tau$. The numerical observers employed in this study are described below.

\subsubsection{Ideal observer and ResNet-based observer}
The ideal observer (IO) is an observer that utilizes all available statistical information about the task at hand in order to maximize task performance. An IO test statistic $t_{\rm IO}(\hat{\f})$ is any monotonic function of the likelihood ratio \cite{barrett}:
\begin{equation}
\Lambda(\hat{\f})=\frac{p(\hat{\f}\mid{H_{1}})}{p(\hat{\f}\mid{H_{0}})},\\
\end{equation}
where $p(\hat{\f}\mid{H_{0}})$ and $p(\hat{\f}\mid{H_{1}})$ are the conditional probability density functions that describe image estimate $\hat{\f}$ under hypothesis $H_0$ and $H_1$. The exact computation of an IO test statistic based on $\Lambda(\hat{\f})$ is intractable in general, and Markov-chain Monte Carlo (MCMC) techniques have been proposed to approximate it \cite{mcmc, gan-mcmc}. Recently, it has been empirically shown that the IO can be approximated by a neural network-based observer \cite{weimin_cnnio}. In this study, a residual neural network-based (ResNet-based) classifier of sufficient capacity trained on a large labeled training dataset was employed in order to approximate the IO. This will henceforth be referred to as the ResNet-IO. Note that if this network does not possess the capacity to accurately approximate $t_{\rm IO}(\hat{\f})$, the resulting NO will be simply referred to as a ResNet-based observer. In this case, the ResNet-based observer is a sub-optimal observer. 


\subsubsection{Hotelling observer (HO) and regularized Hotelling observer (RHO)}\label{sec:bkd-rho}
The Hotelling Observer (HO) is the optimal numerical observer under the condition that the employed test statistic is a linear function of the data\cite{barrett}. The test statistic for the HO is defined as:
\begin{linenomath}
\begin{align}
  t_{\rm{HO}}(\hat{\f}) &= \textbf{w}^\top_\mathrm{HO}\hat{\f},\\
  \intertext{where}
  \mathbf{w}_\mathrm{HO} &= \vec{K}(\hat{\f})^{-1}\Delta{{\bar{\f}}}\\
  \intertext{is known as the Hotelling template and}
  \vec{K}(\hat{\f}) &= \frac{1}{2}\left(\mathbf{K}_0(\hat{\f})+\mathbf{K}_1(\hat{\f})\right).
\end{align}
\end{linenomath}
Here, $\mathbf{K}_0(\hat{\f})$ and $\mathbf{K}_1(\hat{\f})$ denote the covariance matrices of $\hat{\f}$ under the hypotheses $H_0$ and $H_1$, and $\Delta\bar{{\f}} = \mathbb{E}(\hat{\f} \mid H_1) - \mathbb{E}(\hat{\f} \mid H_0)$ is the difference between the condition mean of $\hat{\f}$ under the two hypotheses.

In some cases, the covariance matrix $\mathbf{K}(\hat{\f})$ can be ill-conditioned and therefore its inverse cannot be stably computed. To address this, a regularized Hotelling observer (RHO) can be employed. The singular value decomposition of $\mathbf{K}$ can be written as:
\begin{linenomath}
\begin{align}
    \mathbf{K} = \sum_{i=1}^{R}\sigma_i \mathbf{v}_i\mathbf{u}_i^\dagger,
\end{align}
\end{linenomath}
where $R$ is the rank of $\mathbf{K}$, $\sigma_1 \geq \sigma_2 \geq \dots \geq \sigma_R$ are the singular values of $\mathbf{K}$, $\mathbf{u}_i$ and ${\mathbf{v}}_i$ are the right and left singular vectors respectively, and $^\dagger$ denotes the complex conjugate transpose operation. The truncated pseudoinverse $\vec{K}_\lambda^+$ of $\vec{K}$ can be employed as a stable approximation of $\vec{K}^{-1}$:
\begin{linenomath}
\begin{align}\label{eq:RHO}
    \mathbf{K}_{\lambda}^+ = \sum_{i=1}^{P}\frac{1}{\sigma_i} \mathbf{u}_i\mathbf{v}_i^\dagger,
\end{align}
\end{linenomath}
where $\lambda$ is a threshold for sigular value and $P$ is chosen to satisfy $\sigma_P \geq \lambda\sigma_1 > \sigma_{P+1}$. The truncated pseudoinverse can then be used to construct the regularized Hotelling (RHO) template, which is then used to obtain the RHO test statistic:
\begin{linenomath}
\begin{align}
  t_{\rm RHO}(\hat{\f}) = \mathbf{w}_{\rm RHO}(\lambda)^\top \hat{\f} = \left(\mathbf{K}_{\lambda}^+\Delta{{\bar{\f}}}\right)^\top \hat{\f}.
\end{align}
\end{linenomath}
\subsubsection{Gabor channelized Hotelling observer (Gabor CHO)}
To compute a channelized Hotelling observer (CHO) template, the image data $\hat{\f}$ is first transformed into a vector $\vec{v} \in \mathbb{E}^q, ~ q < n$, known as the {channel output}, via a transformation $\vec{v} = \vec{T}\hat{\f}$, where $\vec{T} \in \mathbb{E}^{q\times n}$ is known as the channel matrix. The test statistic of CHO is then computed as
\begin{align}
    t_{\rm CHO}(\hat{\f})=\vec{w}^\top_\mathrm{CHO}\vec{v},
\end{align}
where $\vec{w}_\mathrm{CHO}=\vec{K}_{\vec{v}}^{-1}\Delta{\mathbf{\bar{\vec{v}}}}$ and $\vec{K}_{\vec{v}} = \frac{1}{2}(\mathbf{K}_{\vec{v},0} + \mathbf{K}_{\vec{v}, 1})$ is the covariance matrix of the channelized image data. Here, $\vec{K}_{\vec{v},0}$ and $\vec{K}_{\vec{v},1}$ denote the covariance matrices of $\vec{v}$ under the two hypotheses $H_0$ and $H_1$. The CHO with Gabor channels (Gabor CHO) can be considered as an anthropomorphic observer\cite{barrett, gabor-zhang2006effect, gabor-eckstein, gabor-yu2013}. The channel matrix $\vec{T}$ employed in the Gabor CHO is specified as follows. A Gabor function $C_i$ corresponding to the $i$th row of $\vec{T}$ is defined in the spatial domain by multiplying a sinusoidal wave with a Gaussian function:
\begin{linenomath}
\begin{align}
    C_i(x,y)=\exp{\left(-(4\ln{2})\frac{x^2+y^2}{w_i^2}\right)}\cos{[2\pi\nu_i(x\cos{\theta_i+y\sin{\theta_i}})+\phi_i]},
\end{align}
\end{linenomath}
where $w_i$ is the channel width, $\nu_i$ is the central frequency, $\theta_i$ is the orientation and $\phi_i$ is the phase. The element $v_i$ of the channel vector $\vec{v} = \vec{T}\hat{\f}$ is then given by the scalar product of the discretized version of $C_i$ with the 2D image representation of $\hat{\f}$.

\subsection{Deep learning-based image super-resolution}
In the context of a image super-resolution problem, a LR image $\f_{\rm LR} \in \mathbb{E}^{n'}, n' \in \mathbb{N}, n' \leq n$, can be formally thought of as being related to the sought-after HR image $\f_{\rm HR} \in \mathbb{E}^n$ via the equation:
\begin{linenomath}
\begin{align}
    \f_{\rm LR} = \vec{H}_{\rm blur}\f_{\rm HR} + \vec{n},
\end{align}
\end{linenomath} 
\noindent where $\vec{H}_{\rm{blur}} \in \mathbb{E}^{n'\times n}$ represents a degradation operator that removes the higher spatial frequencies from $\f_{\rm HR}$ and $\vec{n}$ denotes the noise.
Given a specific LR image, an estimate $\f_{\rm SR} \in \mathbb{E}^n$ of the original high-resolution (HR) image can be obtained by use of image super-resolution methods. However, this is a challenging ill-posed inverse problem. 
In recent years, deep learning has been widely applied to achieve image super-resolution\cite{dl-sisr-review, srcnn,lai2018fast,srgan}. A popular class of deep learning-based approaches calls for establishing a mapping from the space of LR images to the space of HR images:
\begin{linenomath}
\begin{align}
    \f_{\rm SR} = S_{\bm\theta} (\f_{\rm LR}),
\end{align}
\end{linenomath}
where $S_{\bm\theta}$ is a deep neural network parametrized by $\bm\theta$. For several supervised learning approaches, a training dataset of size $D$ consisting of paired low and high-resolution images, $\left\{ (\f_{\rm LR}^{(i)}, \f_{\rm HR}^{(i)}) \right\}_{i=1}^D$, is utilized. A loss function is constructed based on a distance metric $\mathcal{L}(S_{\bm\theta}(\f_{\rm LR}^{(i)}), \f_{\rm HR}^{(i)})$ between a super-resolved (SR) image and an HR image, 
and the optimal parameters $\hat{\bm\theta}$ are estimated by approximately minimizing the loss function over the dataset:
\begin{linenomath}
\begin{align}
    \hat{\bm\theta} = \argmin_{\bm\theta} \frac{1}{D}\sum_{i=1}^D \mathcal{L}\left(S_{\bm\theta}(\f_{\rm LR}^{(i)}), \f_{\rm HR}^{(i)}\right).
\end{align}
\end{linenomath}\par
Various loss functions such as $\ell_1$ or $\ell_2$ loss, or a perceptual loss \cite{perceptual}, can be used to define $\mathcal{L}$. Additionally, an adversarial loss that attempts to match the distribution of SR images to the distribution of original HR images can also be employed \cite{srgan}. The two DL-SR networks considered in this study are the super-resolution convolutional neural network (SRCNN)\cite{srcnn} and the super-resolution generative adversarial network (SRGAN)\cite{srgan}. \par

\noindent\begin{minipage}{0.35\linewidth}
\hphantom{In} The architectures of these two networks are shown in \autoref{fig:srnetworks}. The architecture of the SRCNN consists of feed-forward convolutional layers interspersed with pointwise rectified linear unit (ReLU) nonlinearities \cite{srcnn, relu}. The SRGAN architecture consists of a generative network, which is an image-to-image mapping network consisting of convolutional residual blocks interspersed with pointwise ReLU nonlinearities. A discriminator network is jointly trained along with the generative network, and provides the adversarial loss for matching the distribution of generated SR images to the distribution of HR images\cite{srgan}.
\end{minipage}
\indent\begin{minipage}{0.6\linewidth}
\centering
 \includegraphics[width=\textwidth]{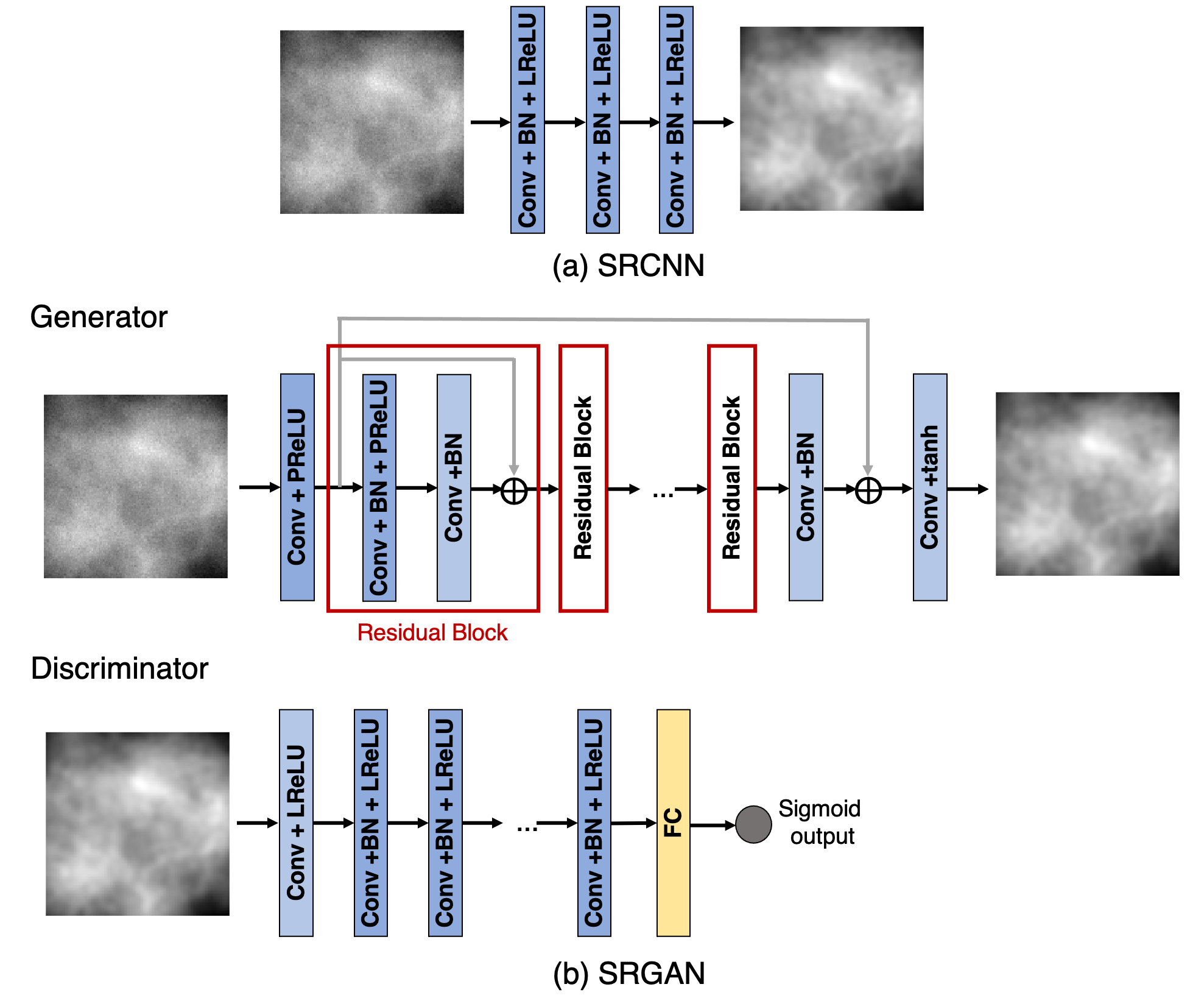}
    \captionof{figure}{Architecture of the super-resolution networks employed in our study, including (a) SRCNN and (b) SRGAN.}
    \label{fig:srnetworks}
\end{minipage}\par

\section{Numerical Studies}\label{sec:numerical}
Computer-simulation studies were employed to objectively evaluate the DL-SR methods described above with two binary signal detection tasks -- (i) a Rayleigh detection task and (ii) a microcalcification (MC) cluster detection task. The NOs described in \autoref{NOs review} were computed on the super-resolved images, as well as the LR and the true HR images, in order to objectively assess the impact of DL-SR on the considered tasks. 

\subsection{Clustered lumpy background (CLB)}\label{sec:clb}
The CLB model was developed by Bochud et al. \cite{clb} for generating random backgrounds that resemble mammographic textures. The value of a CLB image at position $\textbf{r}$ is:
 \begin{linenomath}
\begin{align}\label{eqn:clb}
\f_{b}(\textbf{r})=\sum_{k=1}^{K} \sum_{n=1}^{N_k}l(\textbf{r}-\textbf{r}_k-\textbf{r}_{kn}, \textbf{R}_{\theta_{kn}}), ~\text{ where }~ l(\textbf{r},\textbf{R}_{\theta})=\exp\Big({-\alpha\frac{{\|{{\textbf{R}_{\theta}\textbf{r}}}\|}^\beta}{L(\textbf{R}_{\theta}\textbf{r})}}\Big).
\end{align}
\end{linenomath}
Here, $l(\textbf{r},\textbf{R}_{\theta})$ is known as the {blob function}. The integer $K$ denotes the number of clusters that was sampled from a Poisson distribution with a mean of $\bar{K}:K \sim \mathrm{Poiss}(\bar{K})$, $N_k$ specifies the number of blobs in the $k^{\rm th}$ cluster sampled from a Poisson distribution with the mean of $\bar{N}:N\sim \mathrm{Poiss}(\bar{N})$, $\textbf{r}_k$ indicates the center location of the $k^{\rm th}$ cluster sampled uniformly over the field of view, and $\textbf{r}_{kn}$ represents the center location of the $n^{\rm th}$ blob in the $k^{\rm th}$ cluster sampled from a Gaussian distribution with the center of $\textbf{r}_k$ and standard deviation of $\sigma$. The matrix $\textbf{R}_{\theta_{kn}}$ represents the rotation corresponding to the angle $\theta_{kn}$ sampled from a uniform distribution between 0 and 2$\pi$,
$L(\textbf{r})$ refers to the radius of the ellipse with half-axes $L_x$ and $L_y$, and $\alpha$ and $\beta$ are adjustable coefficients. The parameters of the CLB model employed in both the Rayleigh detection task and MC cluster detection task are shown in \autoref{clb model}. 
\begin{table}[htp!]
    \centering
    \begin{tabular}{@{}lllllll@{}}
    \toprule
    $\bar{K}$&$\bar{N}$ &$L_x$ & $L_y$ & $\alpha$ &$\beta$ & $\sigma$ \\
    150 & 20 & 5 & 2 & 2.1 & 0.5 & 12 \\ \bottomrule
    \end{tabular}
    \vspace{5pt}
    \caption{Parameters for generating CLB images}
    \label{clb model}
\end{table}
\subsection{Rayleigh detection task with a clustered lumpy background model}
The Rayleigh detection task is a natural task for assessing the resolution properties of imaging  systems, and has been employed previously for optimizing tomographic imaging systems \cite{rayleigh, rayleigh_ct}. This is a binary signal detection task, in which the hypothesis $H_0$ corresponds to a signal $\f_{s_0}$ consisting of two adjacent point objects, and the hypothesis $H_1$ corresponds to a signal $\f_{s_1}$ consisting of a single line object.

\subsubsection{Simulated image data for Rayleigh detection task}\label{rayleigh dataset}
Given the definition of signals $\f_{s_0}$ and $\f_{s_1}$ provided above, the generation of low-resolution images under $H_0$ and $H_1$ can be written as:
\begin{linenomath}
\begin{align}\label{eqn:rayleigh}
    H_0 ~ &: ~ \mathbf{f}_{\text{LR}} = \vec{H}_{\text{blur}}\mathbf{f}_0 + \vec{n} \equiv \vec{H}_{\rm{blur}}(\mathbf{f}_b + \mathbf{f}_{s_0}) + \vec{n} \\
    H_1 ~ &: ~ \mathbf{f}_{\text{LR}} = \vec{H}_{\text{blur}}\mathbf{f}_1 + \vec{n} \equiv \vec{H}_{\rm{blur}}(\mathbf{f}_b + \mathbf{f}_{s_1}) + \vec{n},
\vspace{-0.5cm}
\end{align}
\end{linenomath}
where $\mathbf{f}_b$ denotes a CLB image of size $128\times 128$ with parameters defined in \autoref{clb model}, and $\vec{n}$ denotes {\parfillskip=0pt \emergencystretch=.5\textwidth \par}

\vspace{3pt}
\noindent\begin{minipage}{0.37\linewidth}
the measurement noise. Given an adjustable parameter $L$, termed as the {signal length}, $\mathbf{f}_{s_0}$ is specified by first defining two Kronecker delta functions separated by a distance of $L-2$, and convolving them with a Gaussian function of standard deviation 1.375 pixels. The signal $\mathbf{f}_{s_1}$ is specified by first defining a horizontal line of length $L$, which is subsequently convolved with the same Gaussian function. The signals are inserted such that the centers of the  
\end{minipage}
\hfill\begin{minipage}{0.6\linewidth}
\captionsetup{type=figure}
\includegraphics[width=\textwidth]{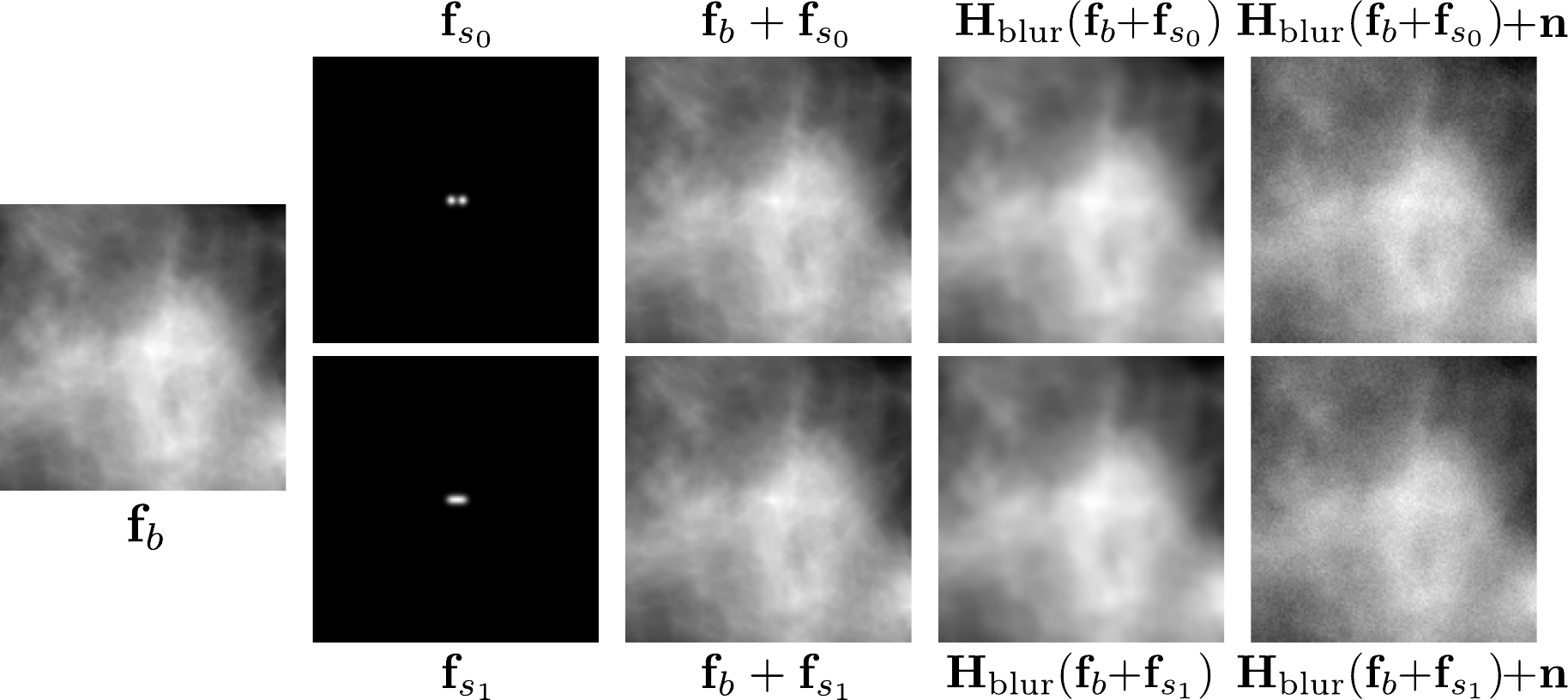}
\vspace{1pt}
\captionof{figure}{Clustered lumpy background ($\mathbf{f}_b$), signals ($\mathbf{f}_{s0}$, $\mathbf{f}_{s1}$), and combined images of the Rayleigh detection task.}
\label{fig:rayleigh_setup}
\end{minipage}
\medskip

\noindent signals coincide with the center of the image. The Rayleigh detection task was performed independently on the following datasets, where the HR dataset consisting of images of the type

\noindent\begin{minipage}{0.49\linewidth}
\vspace{5pt}
\captionsetup{type=figure}
\includegraphics[width=\textwidth]{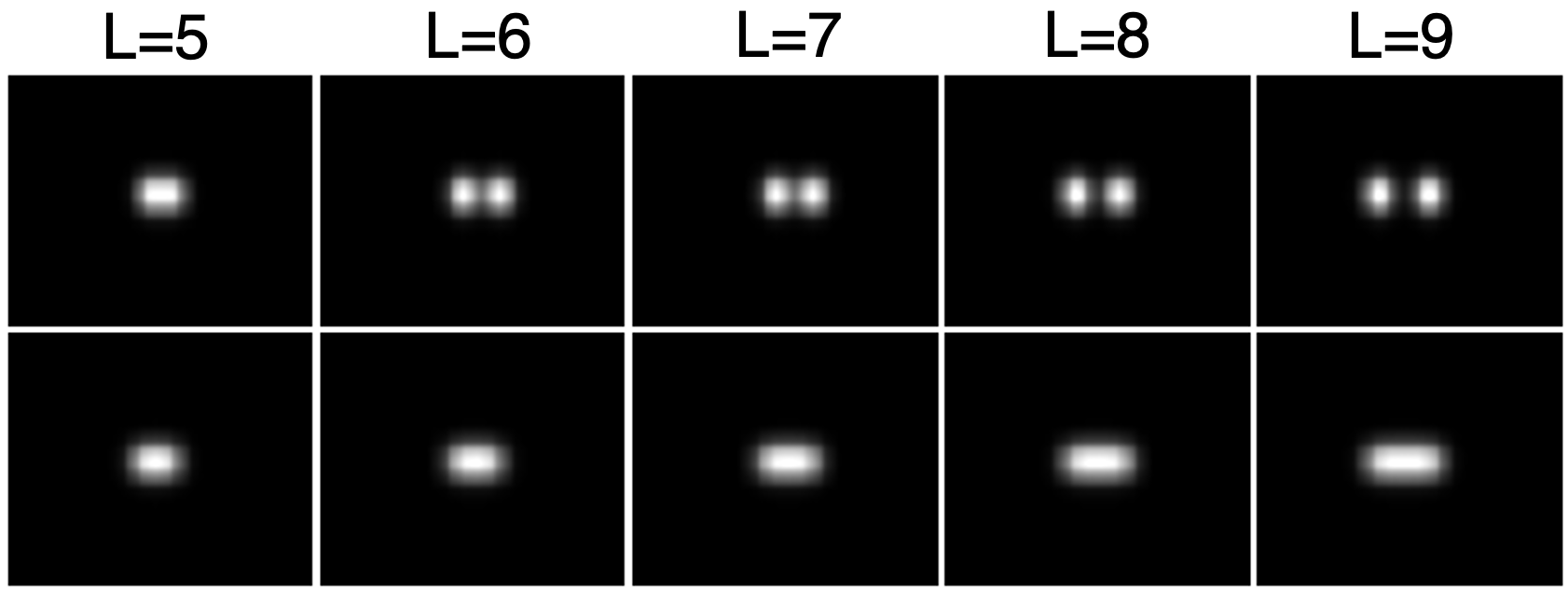}
\vspace{1pt}
\captionof{figure}{Example signal image ROIs with respect to different signal lengths.}
\vspace{7pt}
\label{fig:siglen_sweep_examples}
\end{minipage}
\hfill\begin{minipage}{0.48\linewidth}
 
\begin{align}
\f_{\text{HR}} = \mathbf{f}_i + \vec{n}, ~ i = 0,1,
\end{align}
the LR dataset consisting of images of the type
\begin{align}
\f_{\text{LR}} = \vec{H}_{\text{blur}}\mathbf{f}_i + \vec{n}, ~ i = 0,1,
\end{align} 
and the SR dataset consisting of images of type
\begin{align}
\f_{\text{SR}} = S(\vec{H}_{\text{blur}}\mathbf{f}_i + \vec{n}), ~ i = 0,1, 
\end{align}
\end{minipage}

\vspace{5pt}
\noindent where $\vec{H}_{\text{blur}}$ represents a Gaussian filter with standard deviation of 1.5 pixels, and $\f_i = \f_b + \f_{s_i}$, as defined in \autoref{eqn:rayleigh}. Here, $S$ denotes the DL-SR operation performed by either the SRCNN or the SRGAN, and $\vec{n}$ denotes the sum of pixel-wise independent and identically distributed (i.i.d.) Poisson noise with standard deviation scaled by $\sigma_p=0.013$ and i.i.d. Gaussian noise with standard deviation $\sigma_g=0.35$. The simulation of an example LR image according to the described procedure is shown in \autoref{fig:rayleigh_setup}.

Two separate studies were formulated based on the Rayleigh detection task:
\begin{enumerate}
    \item {Signal length variation study:} In this study, the signal length parameter $L$, which pertains to the distance between the two point objects in $\mathbf{f}_{s_0} $ or the length of the line in $\mathbf{f}_{s_1}$, was varied to investigate the resolving power of the DL-SR algorithms. The signal lengths of $L=\{5,6,7,8,9\}$ were employed in this study as shown in \autoref{fig:siglen_sweep_examples}.
    
    \item {Network complexity variation study:} In order to investigate how the DL-SR network complexity correlates with the task-performance for a fixed object model and task design, a network complexity variation study was conducted where the number of layers of a DL-SR network was varied. The SRGAN employs an additional tunable parameter controlling the trade-off between the MSE loss and the discriminative loss whose optimal value may depend, among other factors, on the number of layers in the network. Hence, only SRCNN was employed in this study. 
\end{enumerate}

\begin{figure}[ht]
\includegraphics[width=\textwidth]{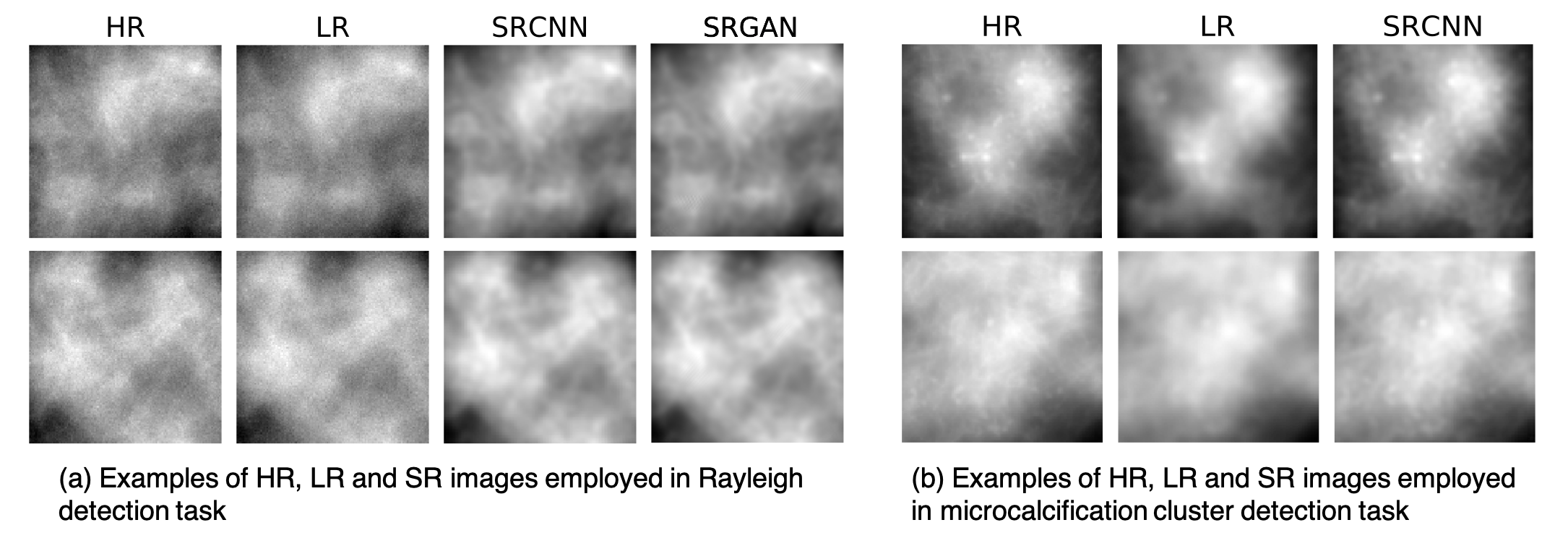}
\caption{Examples of (a) HR, LR and SR images from SRCNN and SRGAN in Rayleigh detection task and (b) HR, LR and SR images from SRCNN in MC cluster detection task.}
\label{fig:rayleigh_mc_examples}
\end{figure}

\subsubsection{Training details for the DL-SR networks}
For the signal length variation study, both the SRCNN and SRGAN were trained and evaluated. The training and validation data for SRCNN consisted of 5,000 and 625 class-balanced signal present/absent images respectively. For SRGAN training, Due to more trainable parameters in the SRGAN, 20,000 images were used for training and 2,000 images were used for validation, respectively, Examples of HR, LR and SR images produced by the networks are shown in~\autoref{fig:rayleigh_mc_examples}(a).

For the architecture variation study, seven SRCNNs with varying number of convolutional layers ranging from 2 to 8 were employed. For all the SRCNNs, the filter size in the first layer was fixed to $9\times9$ whereas the filter size for the other layers was fixed to $5\times5$.
The number of filters in all the layers was fixed to 32, except the last layer, where the number of filters was fixed to 1. All SRCNNs were trained on 15,000 images and validated on 3,000 images with class balance.

The SRCNN was trained with an MSE loss and the SRGAN was trained by use of an MSE loss and an adversarial loss. All DL-SR networks to be evaluated in the Rayleigh detection task were trained on mini-batches at each iteration by use of the Adam optimizer\cite{adam}. The DL-SR models that achieved the best performance on the validation set were used for evaluation. Both DL-SR networks were implemented under the TensorFlow 2.0 framework and trained on NVIDIA GPUs.\par

\subsection{Microcalcification cluster detection task with a clustered lumpy background model}

Motivated by the clinical value of detecting microcalcification (MC) clusters in mammograms that may be associated with malignancy in breast lesions \cite{mc-clinical-value, mc-cluster-relevance}, a {stylized} SKS/BKS binary signal detection task of identifying an image with or without a MC cluster present was studied.
The objective of this study was to determine how the capacity of a numerical observer affects observer performance on super-resolved images. In essence, whether or not SR aids the performance of sub-optimal observers was systematically studied.

\subsubsection{Simulated image data for MC cluster detection task}
The HR MC cluster dataset was created as follows. First, 128$\times$128 CLB images were created to simulate the mammographic backgrounds, as described in \autoref{sec:clb}. The signal-absent HR images $\f_0$ correspond to the case where $\f_{s_0} = 0$ and hence, were kept equal to the CLB images. The signal insertion pipeline employed to generate the signal-present HR image $\f_1$ is described as follows. A set of eleven $200\times200$ MC clusters segmented from digital mammograms acquired with the Selenia Dimensions system (Hologic, Inc.), available at \url{https://github.com/LAVI-USP/MCInsertionPackage}\cite{MC_dataset}, were employed to model the MC cluster signal. First, one out of the eleven segmented MC clusters was chosen at random and a random rotation between 0\degree to 360\degree with zero padding was applied. Next, this rotated image $\mathbf{s_{MC}}$ was cropped to size of $128\times128$ and inserted into a CLB $\f_b$ as \cite{mc_insertion}:
\begin{linenomath}
\begin{align}
    \mathbf{f}_1 = \f_b(c\mathbf{s_{MC}} + 1).
\end{align}
\end{linenomath}
The scalar $c$ represents a contrast factor uniformly sampled from the range $\left[0.05, 0.06\right]$ that is chosen to visually match the contrast of real lesion. 

Given the generated HR image, the corresponding LR image was simulated as follows, based on the degradation model described in You, \textit{et al.}\cite{gancircle}.
\begin{equation}
\mathbf{f}_{\text{LR}} = \vec{H}_{\rm blur}\mathbf{f}_i + \vec{n}, \quad i = 0,1.
\end{equation}
Here, $\vec{H}_{\rm{blur}}$ represents a Gaussian blurring operation with standard deviation of 1.5 pixel, followed by a downsampling by a factor of 2. 
Pixel-wise i.i.d. Poisson noise with standard deviation scaled by a factor $\sigma_p=0.0001$, and i.i.d. Gaussian noise with standard deviation $\sigma_g=0.001$ was added to both the HR and LR images. In order to enable direct comparison with the HR and SR images, an additional operation $\vec{U}$ representing upsampling by a factor of 2 was used on the LR images. Similar to the Rayleigh detection task, the MC cluster detection task was performed on the following datasets - (1) the HR dataset consisting of images of the type $\mathbf{f}_{\text{HR}} = \mathbf{f}_i + \vec{n}, ~ i = 0,1$ is one of the MC cluster-absent/present hypothesis, (2) the LR dataset consisting of images of the type $\mathbf{f}_{\text{LR}} = \vec{H}_{\text{blur}}\mathbf{f}_i + \vec{n},  ~ i = 0,1$ along with the additional upsampling operation $\vec{U}$ acting on $\mathbf{f}_{\text{LR}}$, and (3) the SR datasets consisting of $\mathbf{f}_{\text{SR}} = S(\vec{U}\mathbf{f}_{\text{LR}})$, where $S$ denotes the DL-SR operation performed by SRCNN.
\subsubsection{Training details for DL-SR networks}
The SRCNN employed in this study was trained on a dataset of 40,000 images and validated on a dataset of 4,000 images, both with balanced classes. The network was trained with the Adam optimizer \cite{adam} with a learning rate of $5\times10^{-5}$ for 1,000 epochs to minimize the MSE loss. The SRCNN model with the best validation performance was used. Examples of the SR images produced by the SRCNN along with the HR and the LR images are shown in \autoref{fig:rayleigh_mc_examples}(b).

\subsection{Objective evaluation of deep learning-based image super-resolution networks}
\subsubsection{Objective evaluation metrics for the Rayleigh detection task}
\noindent\begin{minipage}{0.55\linewidth}
To evaluate the DL-SR networks with task-based metrics, three NOs, namely the RHO, Gabor CHO and ResNet-IO were employed. The test statistics for the three NOs were computed on the HR, LR and SR images that were centrally cropped to a size of $64\times64$. Receiver operating characteristic (ROC) curves were computed, and the area under the ROC curve (AUC) was employed as a figure-of-merit. All evaluation metrics were computed on balanced test dataset of 40,000 images. Non-parametric estimation of the AUC confidence intervals was carried out using the DeLong's algorithm \cite{delong, delong_alg}, with the help of the pROC package in R \cite{proc}. Additionally, traditional IQ metrics such as PSNR and SSIM were computed on the LR and SR images.\par
\end{minipage}
\indent\begin{minipage}{0.4\linewidth}
    \centering
    \includegraphics[width=\textwidth]{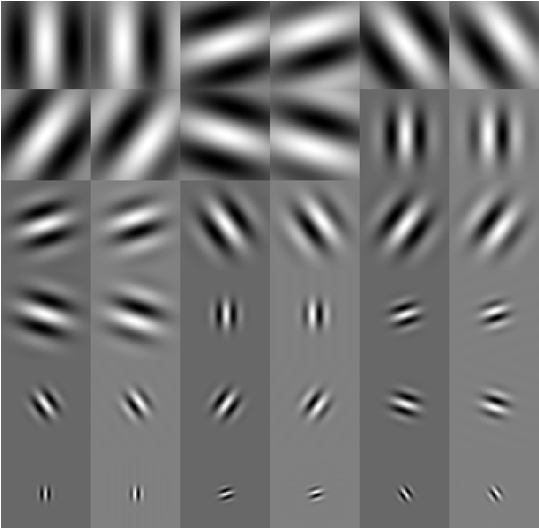}
    \captionof{figure}{Examples of Gabor channel templates}
    \label{fig:gabor channel}
\end{minipage}\par

\vspace{3pt}
To compute the RHO test statistic, 500,000 images containing two point objects and 500,000 images containing the line-shaped object were utilized to estimate the empirical covariance matrix $\vec{K}(\hat{\f})$. The threshold parameter $\lambda$ in \autoref{eq:RHO} was swept in a from $10^{-9}$ to $10^{-4}$ and the detection performance was evaluated on a validation set of 4,000 class-balanced images. The value of $\lambda$ that yielded the best RHO performance on validation data was selected. This RHO with the selected parameter $\lambda$ was applied to a test set consisting of 40,000 class-balanced images.\par

The channel matrix corresponding to the Gabor CHO comprised of a set of 60 Gabor channels. Each Gabor channel was associated with one out of six passbands, one out of five orientations, and one out of two phases. The six passbands each have a spatial frequency bandwidth of 1 octave with center frequency $\nu=3/256,3/128,3/64,3/32,3/16$ and $3/8$ cycles/pixel. The five orientations were $0, 2\pi/5, 4\pi/5, 6\pi/5$ and $8\pi/5$, and the two phases were $0$ and $\pi/2$. Examples of Gabor channel templates are shown in \autoref{fig:gabor channel}. The channelized covariance matrix was estimated using 100,000 images from each class with 500,000 noise realizations for each class.\par

The ResNet-IO, as shown in \autoref{fig:resnetio}(a), was employed to approximate the IO test statistic. In order to obtain a good approximation of the IO using ResNets, the optimum network capacity needs to be determined empirically by sweeping the number of layers used in
\noindent the ResNet architecture and choosing the configuration that gives the best detection performance. A large training dataset must be used in order to correctly represent the data distribution. Here, the network was initialized with

\noindent\begin{minipage}{0.35\linewidth} 
\smallskip
\noindent the help of the RHO template in order to give the best performance and speed up convergence. A family of ResNets comprising of various numbers of residual blocks were trained on a dataset consisting of 100,000 training images and validated on 4,000 images from each of the two classes. The binary cross-entropy loss was minimized by use of Adam optimizer with a learning rate of $1\times10^{-6}$. Additionally, a ``semi-online learning" method in which the measurement noise was generated on-the-fly as described in \cite{weimin_cnnio}
\end{minipage}
\indent\begin{minipage}{0.6\linewidth}
 \centering
    \includegraphics[width=\textwidth]{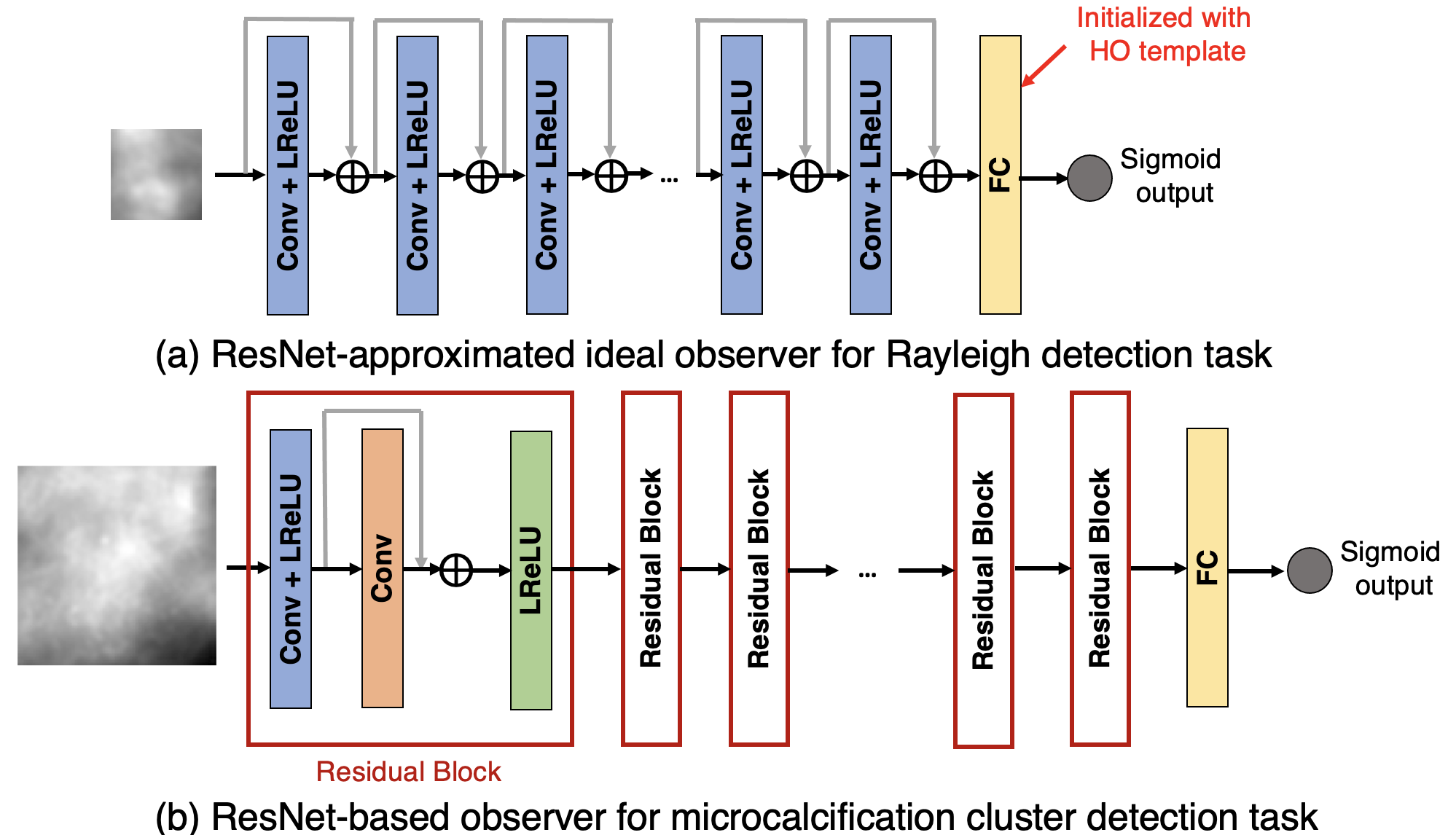}
    \smallskip
    \captionof{figure}{Architectures of (a) ResNet-approximated ideal observer for Rayleigh detection task and (b) ResNet-based observer for microcalcification cluster detection task.}
    \label{fig:resnetio}
\end{minipage}
\smallskip

\noindent was utilized to mitigate the overfitting problem. The ResNet that had the best validation performance was chosen as the ResNet-IO.

\subsubsection{Objective evaluation for the MC cluster detection task} 
As described previously, the objective of this study was to investigate the potential benefit of DL-SR as it relates to the capacity of an NO. A binary signal detection task was conducted to distinguish whether an image contains microcalcification cluster signal or not. In order to assess the task-based performance, a family of ResNet-based observers consisting of 2, 4, 6 or 8 residual blocks respectively were employed in the detection task. The architecture of the ResNet-based observers is shown in \autoref{fig:resnetio}(b).
Each of these observers were trained on class-balanced datasets of sizes 5,000 10,000, 20,000 and 50,000 and 100,000 by minimizing the binary cross-entropy loss, until the detection capability of each observer was fulfilled. Each simulated MC cluster image in the training dataset was augmented four times by flipping. The AUC values produced by the trained ResNet-based observers on a held-out test set including 20,000 images from each class were used to evaluate the signal detection performance. The ResNet-based observer that achieves the best test performance without further improvement with either a deeper network architecture or a larger training dataset could be considered as an approximated IO\cite{weimin_cnnio}.\par

\section{Results}\label{sec:results}
\subsection{Rayleigh task}
\subsubsection{Impact of regularization on the Hotelling observer performance}

\begin{figure}[ht!]
      \centering
    \includegraphics[width=\textwidth]{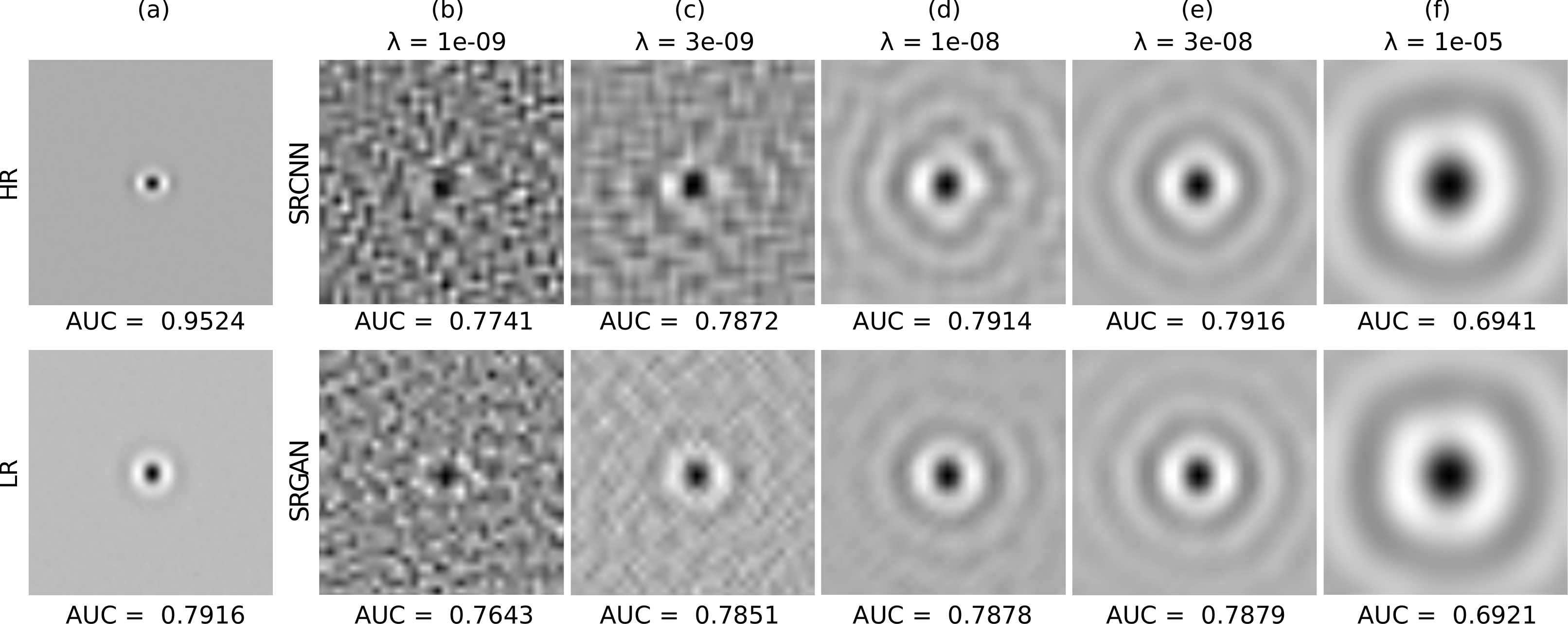}
    \smallskip
    \captionof{figure}{RHO templates computed on (a) HR and LR images, and (b--f) images from SRCNN and SRGAN resulting from sweeping the regularization parameter $\lambda$.}
    \label{fig:rho_template}
\end{figure}

In addition to introducing high frequency features to an LR image, the DL-SR networks also suppress the per-pixel i.i.d. noise added to the LR images. Due to this, the covariance matrix $\mathbf{K}(\hat{\f}_{\rm SR})$ of the super-resolved images is ill-conditioned. Hence, as mentioned in \autoref{sec:bkd-rho}, regularization is needed to stably invert it for obtaining the Hotelling template. Hence, the performance of the RHO depends upon the regularization parameter $\lambda$ employed for truncating the singular values of $\mathbf{K}$. Figure \ref{fig:rho_template} shows the Hotelling templates of the HR images, the LR images and the images super-resolved by the SRCNN and the SRGAN.  It can be seen that for low values of $\lambda$, the Hotelling template is noisy, due to the unstable inversion of $\mathbf{K}$. On the other hand, for high values of $\lambda$, degradation of the signal specificity corresponding to the truncation of singular values can be seen.  

\subsubsection{Impact of signal length on observer performance}
\begin{figure}[ht!]
    \centering
    \includegraphics[width=\textwidth]{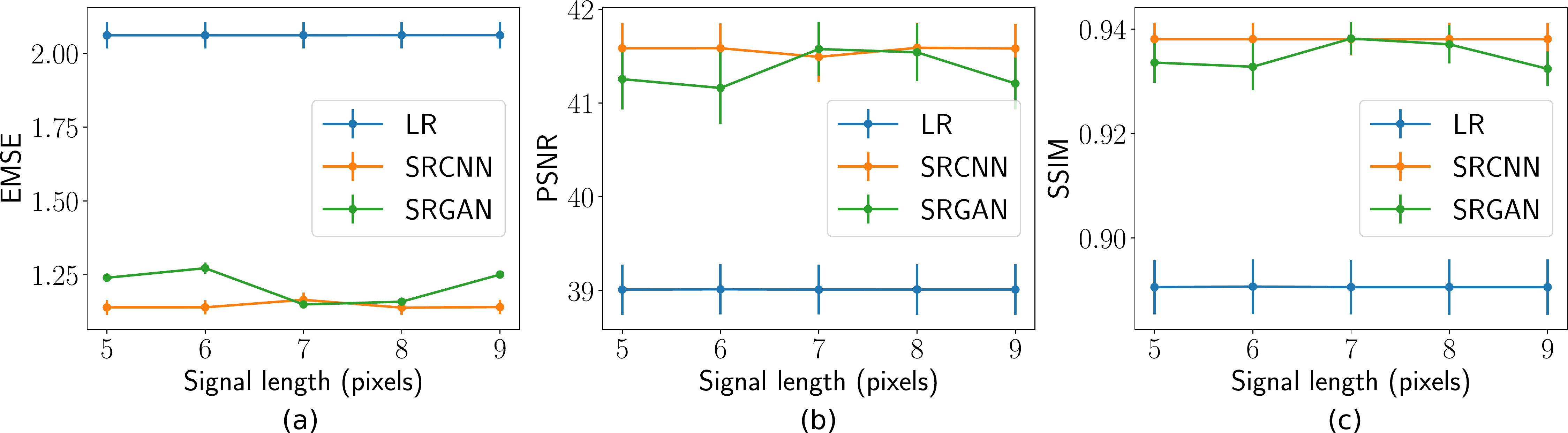}
    \vspace{5pt}
    \caption{Traditional image quality metrics including (a) Ensemble MSE, (b) PSNR, and (c) SSIM of the HR, LR and SR images. Both the SRCNN and SRGAN consistently and significantly improved the image quality across various signal lengths in terms of these traditional metrics.}
    \label{fig:signallengthsweep_trad_metrics}
\end{figure}

\begin{figure}[ht!]
    \includegraphics[width=\textwidth]{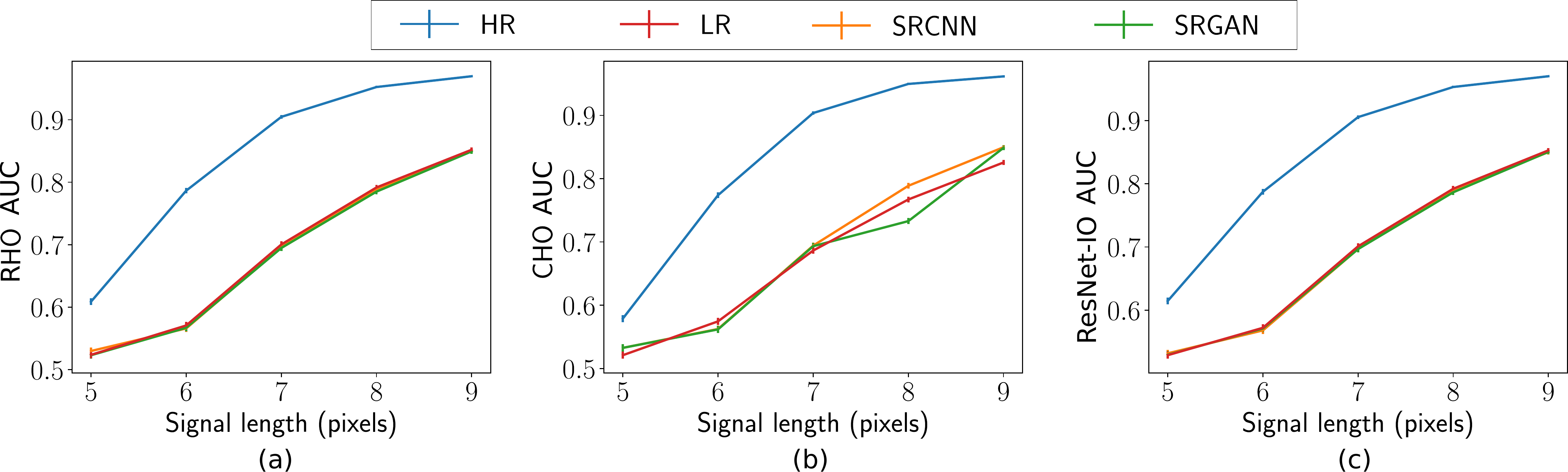}
    \vspace{5pt}
    \caption{AUC values of the (a) RHO, (b) CHO, and (c) ResNet-IO for HR, LR and SR images. It can be seen that the DL-SR resulted in a small improvement in the CHO performance, but no improvement in the RHO and ResNet-IO performance on the LR images. As such, the observer performance on the HR images is much higher than the performance on the LR and SR images.}
    \label{fig:signallengthsweep_aucs}
\end{figure}

The traditional IQ metrics and AUC values for the signal length variation study computed on a class-balanced test set consisting of 40,000 images are plotted in \autoref{fig:signallengthsweep_trad_metrics} and  \autoref{fig:signallengthsweep_aucs} respectively. As seen in \autoref{fig:signallengthsweep_trad_metrics}, the SR images generated by the SRCNN and SRGAN show an improvement in image quality across various signal lengths compared with their LR counterparts in terms of the traditional IQ metrics. Moreover, no significant changes on traditional IQ metrics were observed among SR images when varying signal length. This is due to the fact that the degradation model and DL-SR network architecture were consistent across different signal lengths and the physical difference among images with various signal lengths was minor.\par

However, as shown in \autoref{fig:signallengthsweep_aucs}, DL-SR performance as measured by NO performance provides different insights into the DL-SR behavior. Firstly, it can be seen that AUC values corresponding to all NOs increased consistently along with the increment of the signal length for the HR, LR and both types of SR images. This is due to the fact that with a increasing signal length, the detection task became easier. Secondly, the AUC values corresponding to HR images were significantly greater than those on LR images and SR images. This suggested that the second- and potentially higher-order statistical properties of the images may not be recovered by the DL-SR networks. Thirdly, it is worth noting that in some cases, there was a small improvement in the AUC values of RHO and a small but significant improvement in the AUC values of Gabor CHO corresponding to the SR images as compared to the LR images. This could be interpreted by the fact that both the linear observers, namely the RHO and the Gabor CHO acting on the SR images, have the benefit of having a nonlinear preprocessing block in the form of the DL-SR network. Finally, as shown in \autoref{fig:signallengthsweep_aucs}(c), there was no improvement in the performance of the ResNet-IO as a result of the employed DL-SR networks, which is consistent with the data-processing inequality \cite{dpi}.\par

\subsubsection{Impact of number of layers in DL-SR networks on observer performance}
The traditional IQ metric MSE and the NO performance measured on the LR and SR images as the number of layers in SRCNN was varied are shown in \autoref{fig:layersweep_mse} and \autoref{fig:layersweep}, respectively. As shown in \autoref{fig:layersweep_mse}, the MSEs decreased when the number of layers in SRCNN increased, as expected. This indicates the DL-SR networks improved certain first-order statistics of the images. However, this trend is not always consistent with the NO performance measured by AUC values. As shown in \autoref{fig:layersweep}, it was observed that the AUC values for the RHO measured on SR images were no greater than those computed by use of the LR images. Also, the RHO performance decreased as the number of DL-SR network layers increased. This suggests that the second-order statistical properties of the images were degraded by the DL-SR networks. In order to further analyze this, the singular values of the covariance matrix $\mathbf{K}(\hat{\mathbf{f}}_{\rm SR})$ of the SRCNN-resolved images were computed for networks having different number of layers. As shown in \autoref{fig:rho_svals_archsweep}, the singular values indicate that as the number of layers in the DL-SR network increased, $\mathbf{K}(\hat{\mathbf{f}}_{\rm SR})$ became increasingly ill-conditioned.\par

\noindent\begin{minipage}[t]{0.49\linewidth}
\captionsetup{type=figure}
\includegraphics[width=\textwidth]{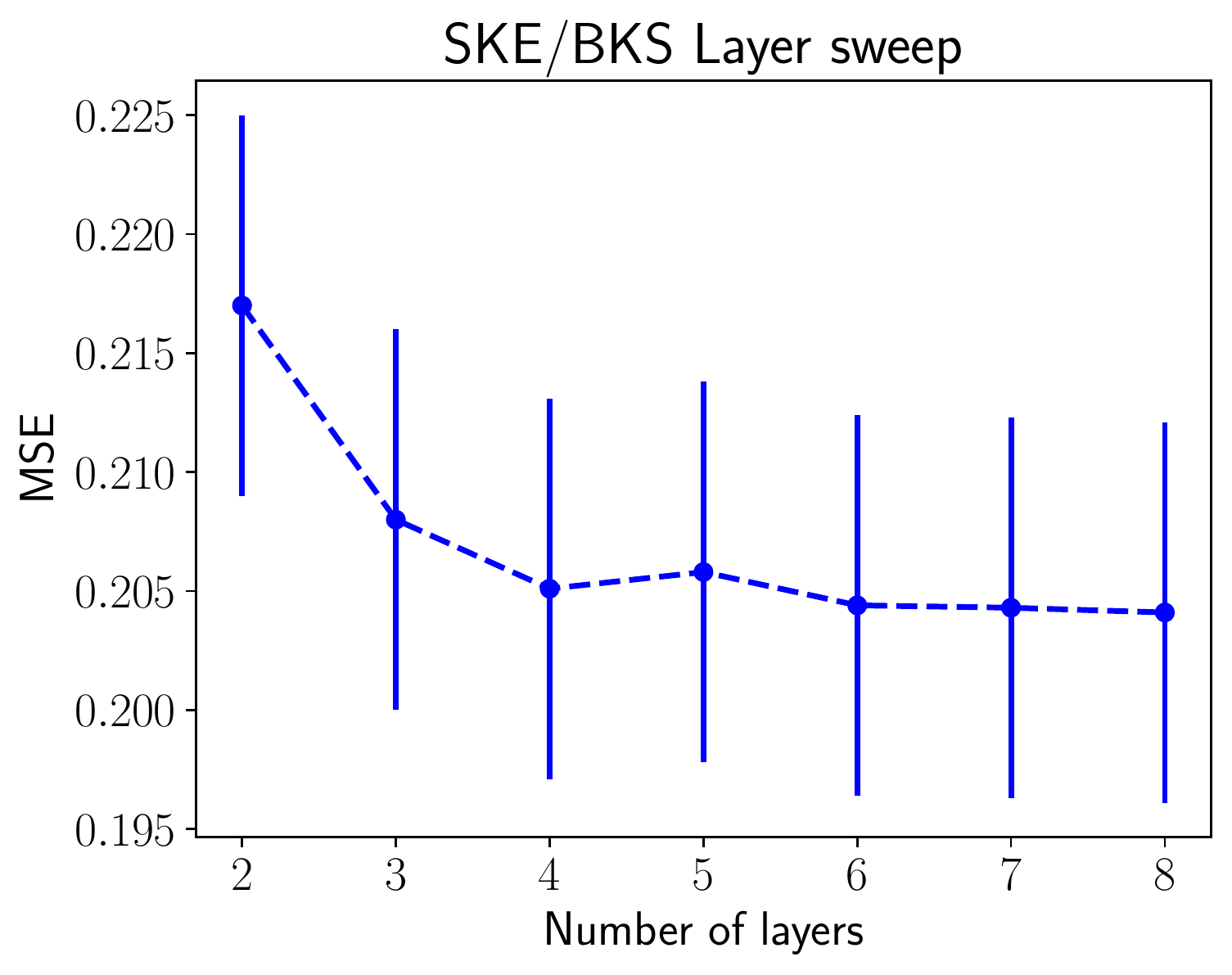}
\captionof{figure}{Ensemble MSE between the SR and the HR images for SR networks with different numbers of layers. The LR images yield an MSE of 0.4369.}
\label{fig:layersweep_mse}
\end{minipage}\quad
\noindent\begin{minipage}[t]{0.49\linewidth}
\captionsetup{type=figure}
\includegraphics[width=\textwidth]{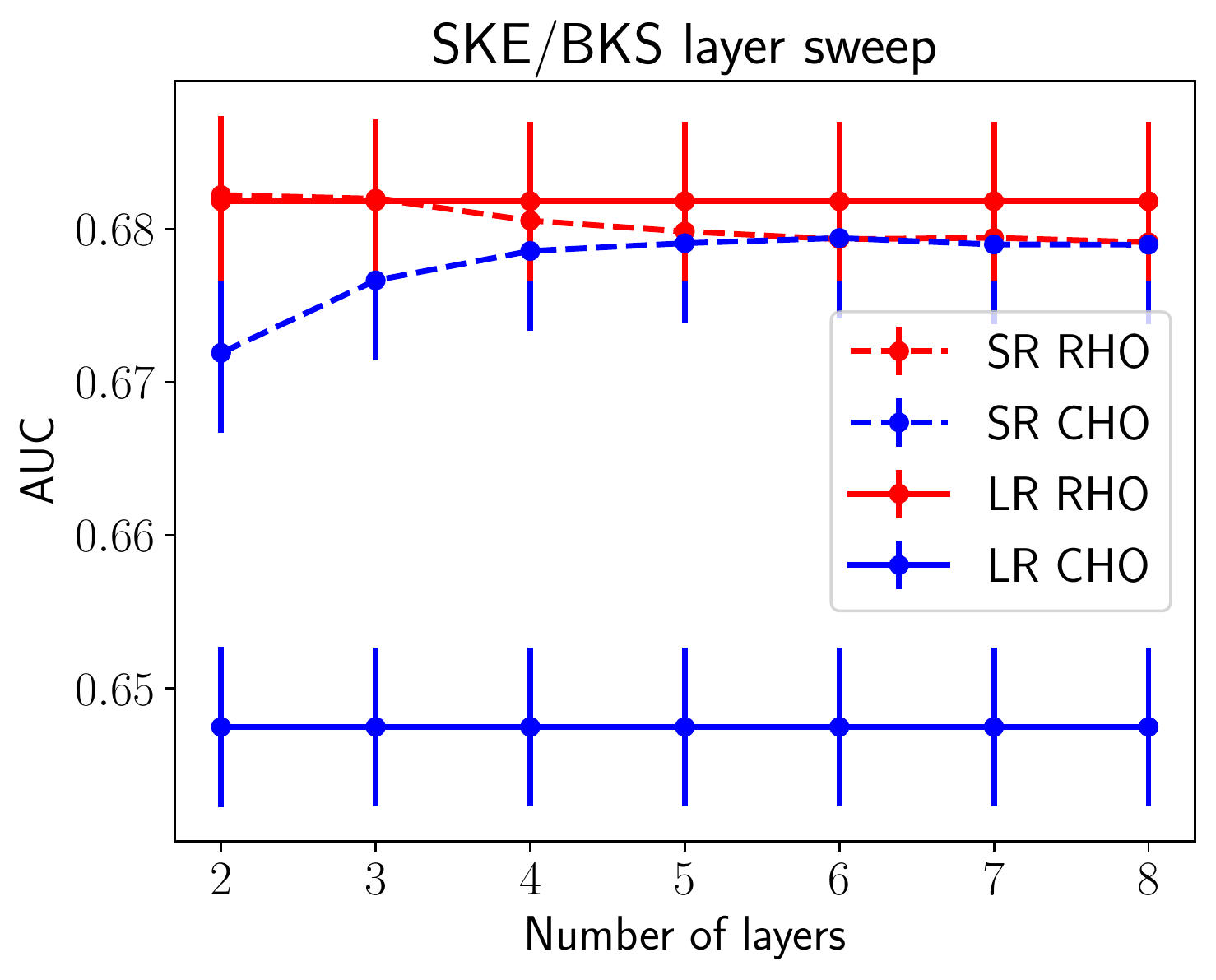}
\captionof{figure}{RHO and CHO performance on SR images and LR images, for SR networks with different numbers of layers.}
\label{fig:layersweep}
\end{minipage}

\noindent\begin{minipage}[t]{0.49\linewidth}
\captionsetup{type=figure}
\includegraphics[width=\textwidth]{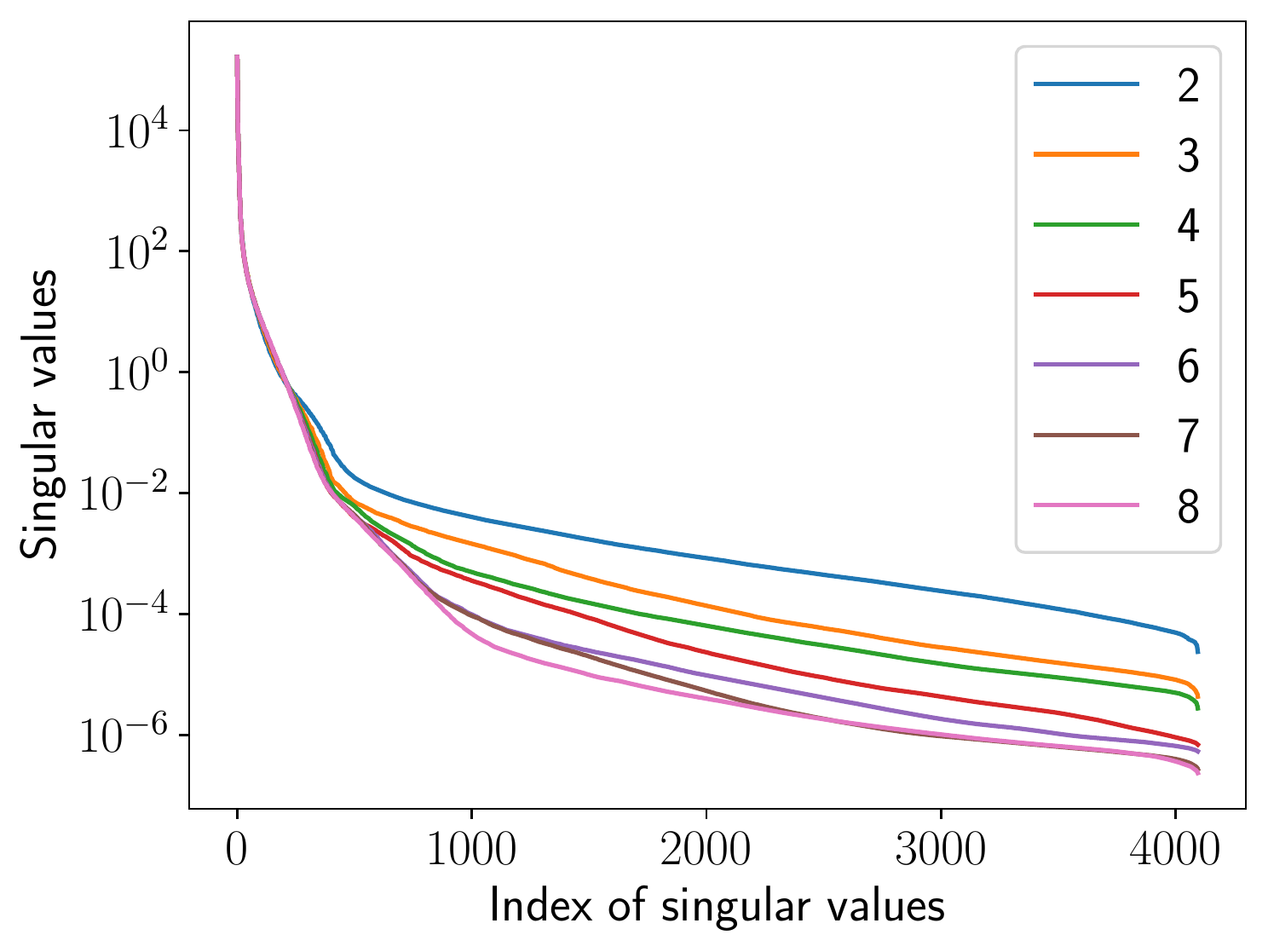}
\captionof{figure}{Singular values of the empirical covariance matrix of the SR images from DL-SR networks of different numbers of layers.}
\label{fig:rho_svals_archsweep}
\end{minipage}\quad
\noindent\begin{minipage}[t]{0.49\linewidth}
\captionsetup{type=figure}
\includegraphics[width=\textwidth]{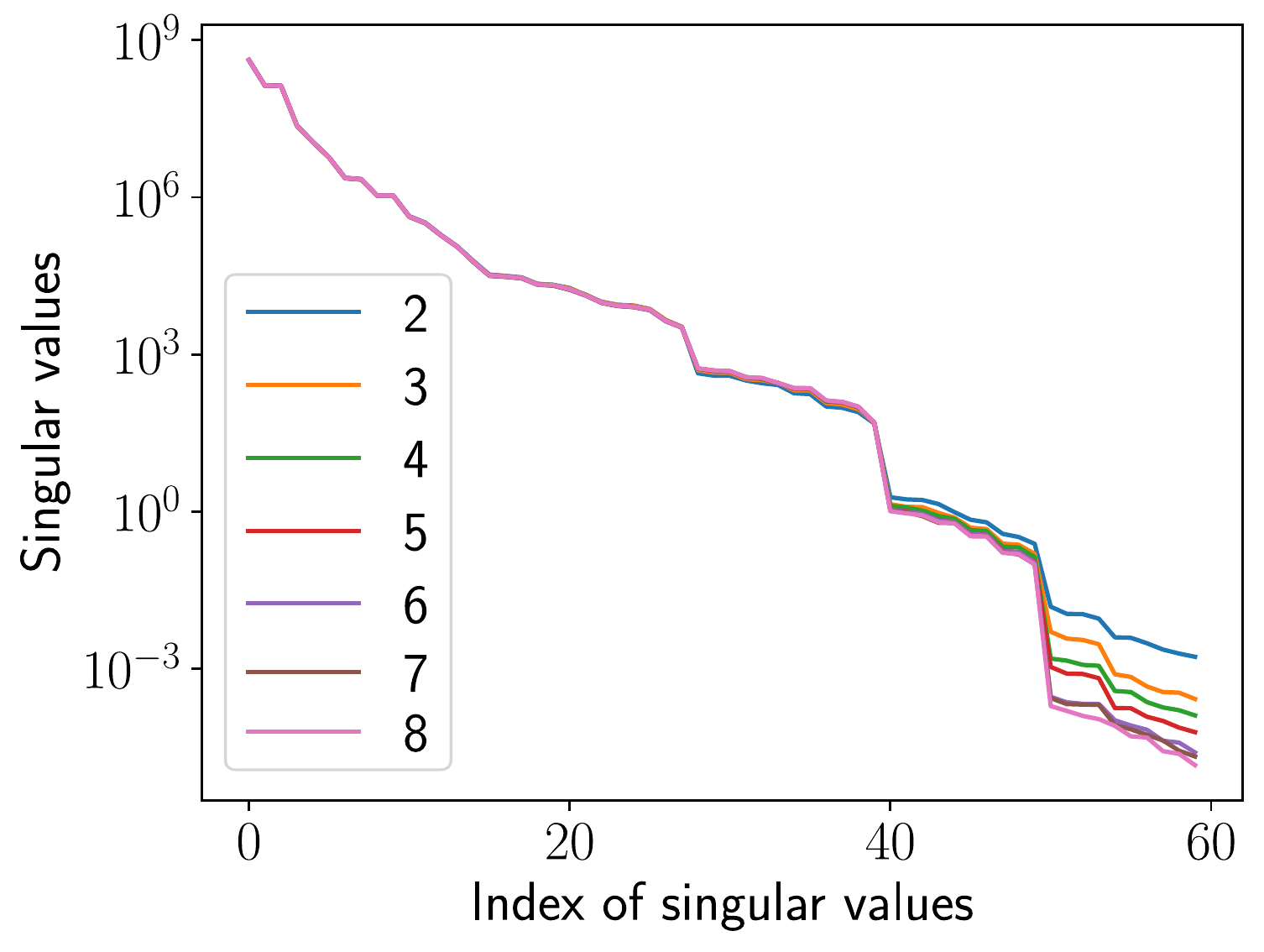}
\captionof{figure}{Singular values of the empirical covariance matrix of the Gabor channelized SR images from DL-SR networks of different numbers of layers.}
\label{fig:cho_svals_archsweep}
\end{minipage}
\medskip

On the other hand, the AUC values for the Gabor CHO on SR images were greater than that measured on LR images, and the performance of Gabor CHO on SR images increased as the number of layer increased from 2 to 6, after which it saturates and reduces slightly for the SRCNN composed of 7 and 8 layers. This suggests that the second-order statistics of the Gabor channelized images were improved by the DL-SR networks, but this improvement reached a plateau as the number of layers increased. The singular values of the covariance matrix $\mathbf{K}_{\mathbf{v}}$ of the Gabor-channelized, SRCNN-resolved images were computed for the DL-SR networks with different numbers of layers. As shown in \autoref{fig:cho_svals_archsweep}, the singular value decay of $\mathbf{K}_{\mathbf{v}}$ is faster for DL-SR networks with more layers, which is similar to the RHO.\par



\subsection{Impact of observer capacity on benefit of DL-SR for MC cluster detection performance}
The objective of this study is to determine how the capacity of a NO relates to its task performance on SR images. The traditional IQ metrics MSE, PSNR and SSIM were computed for the LR and SR images generated by the SRCNN on the MC cluster dataset. As shown in \autoref{tab:mc_metrics}, the image quality measured with these metrics improved for the SRCNN-resolved images compared to the LR counterparts.\par
\begin{table}[htp!]
    \centering
    
    \begin{tabular}{c|c|c|c}
    \hline
         Resolution & Ensemble MSE & PNSR & SSIM\\\hline
         LR & $0.1580\pm0.0104$ & $50.1925\pm0.5390$ & $0.9942\pm0.0006$ \\\hline
         SR & $0.0486\pm0.0021$ & $55.2895\pm0.3546$ & $0.9973\pm0.0002$ \\\hline
    \end{tabular}
    \smallskip
    \caption{Traditional image quality metrics computed on the LR and SR images in the MC cluster detection task.}
    \label{tab:mc_metrics}
\end{table}

The capacity of a ResNet-based observer was varied by varying the number of residual blocks that constitute the ResNet. Figure \ref{fig:auc_vs_nolayers} shows the performance of ResNet-based observers consisting of 2, 4, 6 and 8 residual blocks trained on a dataset of 50,000 images (200,000 considering 4-fold flip-augmentation). It was observed that ResNet-based observers of smaller capacity benefited from the particular DL-SR network employed. In this case, the DL-SR network can be interpreted as an additional pre-preocessing block for the ResNet observer that effectively increases the capacity of the observer. However, as the capacity of the observer was increased, the SR operation gave diminishing returns towards improving the task-performance. As the NO performance plateaued with increasing capacity, it approached ResNet-IO, and the MC cluster detection performance on SR images was no greater than that in LR images. This behavior is consistent with the data processing inequality\cite{dpi}, which suggests that post-processing operations such as image super-resolution will not increase the information content in the image. As a result, the MC cluster detection performance of a ResNet-IO on SR images should not be expected to surpass that of the original LR images.\par
Next, ResNet-based observers of varying depths were trained on datasets consisting of different sizes to fulfill their corresponding capacity for each resolution. For each dataset, the optimal ResNet-based observer was identified based on the best performance on the validation dataset. The results in \autoref{fig:auc_vs_dataset} show the performance of the optimal ResNet-based observer for each dataset size. It was observed that as the amount of available training data increased, the MC cluster detection performance of the ResNet-based observers increased. More interestingly, given a small dataset with limited number of images such as 5,000, 10,000 and 20,000, the DL-SR network indeed improved the detection performance on SR images compared to LR. This demonstrates a situation where the DL-SR operation aided the MC cluster detection performance. For training dataset sizes of 50,000 and beyond, the ResNet-based observer approached the ResNet-IO, and its performance on the images resolved by the DL-SR networks was no better than its performance on the LR images.\par
\vspace{10pt}
\noindent\begin{minipage}[t]{0.49\linewidth}
\captionsetup{type=figure}
\includegraphics[width=\textwidth]{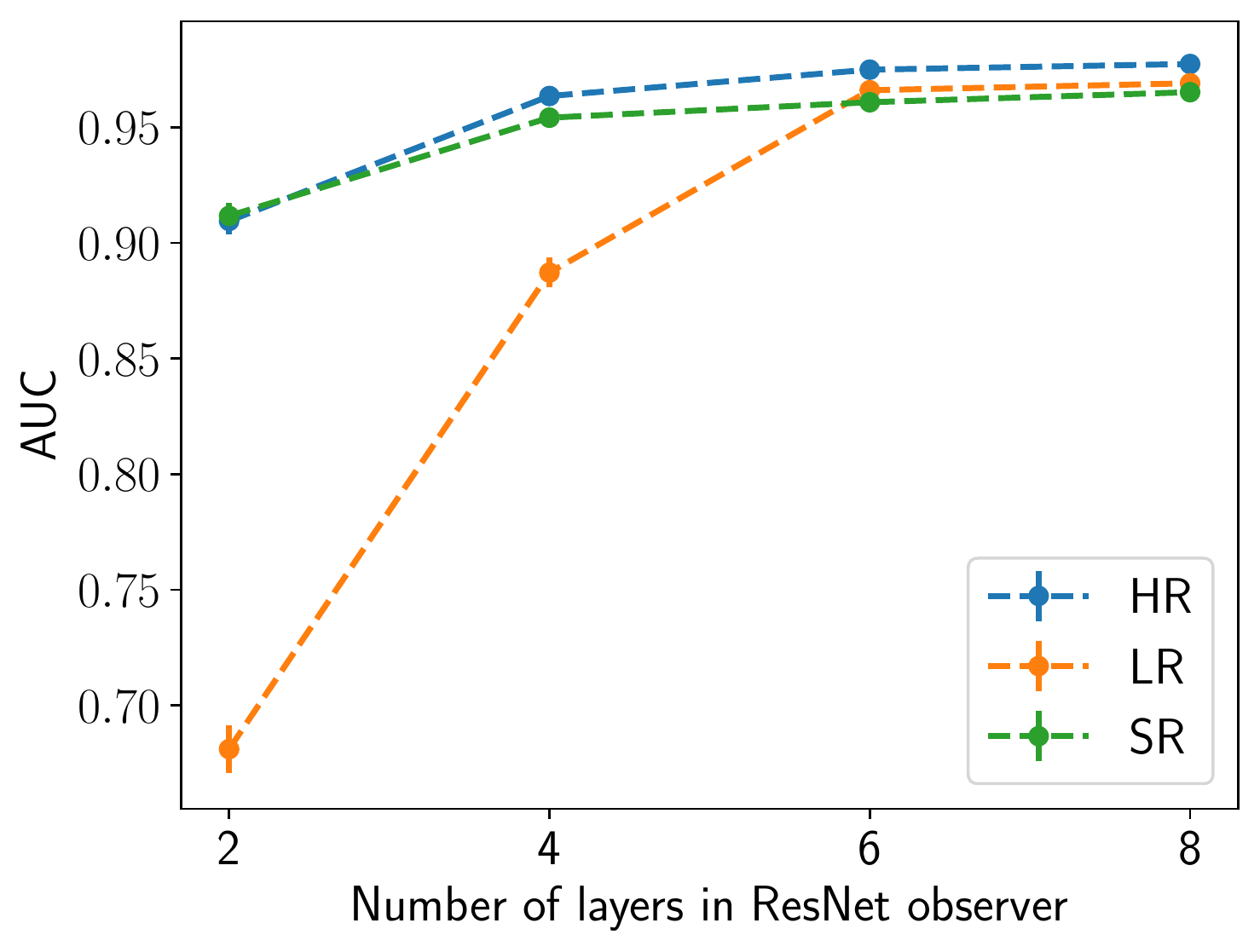}
\captionof{figure}{Performance of ResNet-based observers of different numbers of layers trained on HR, LR, and SR datasets of size 50,000.}
\vspace{20pt}
\label{fig:auc_vs_nolayers}
\end{minipage}\quad
\begin{minipage}[t]{0.49\linewidth}
\captionsetup{type=figure}
\includegraphics[width=\textwidth]{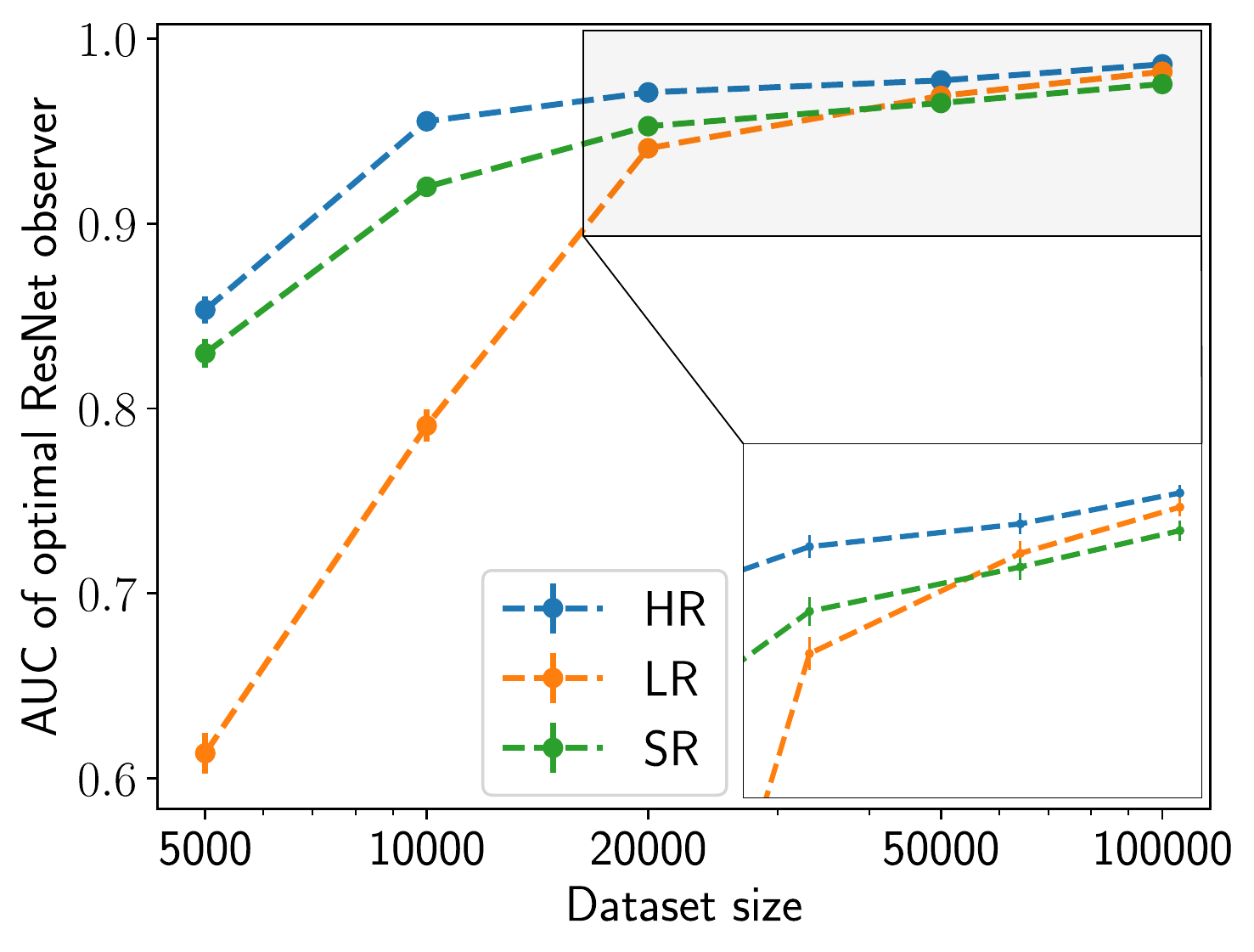}
\captionof{figure}{Performance of the optimal ResNet-based observer for a particular dataset size trained on HR, LR, and SR images.}
\vspace{20pt}
\label{fig:auc_vs_dataset}
\end{minipage}

Both these observations in \autoref{fig:auc_vs_nolayers}  and \autoref{fig:auc_vs_dataset} illustrate that, in the case of sub-optimal neural-network (NN)-based observers, such as those with limited capacity or those trained on limited data, DL-SR networks may be employed to improve the detection performance compared to that achieved on the LR images. However, if the NN-based observer approximates IO, preprocessing the LR images using a DL-SR network will not improve the detection performance of the observer. 


\section{Discussion}\label{sec:discussion}
Deep learning techniques have been adopted for a wide range of medical imaging applications, including image restoration. Despite the fact that different traditional image quality (IQ) metrics have been computed to assess the effect of these deep learning-based methods, a task-based evaluation of these approaches has been largely lacking. A recent study conducted by Li, \textit{et al.} demonstrated that deep neural network (DNN)-based image denoising methods can result in a loss of task-relevant information, despite an improvement in several traditional IQ metrics \cite{kaiyan-denoise}. In a similar vein, this work studies the impact of deep learning-based image super-resolution (DL-SR) on binary signal detection tasks. It is important to reiterate that the main goal of this work is to comprehensively study the impact of DL-SR on task performance for known tasks under known statistical conditions. It is not to explore whether DL-SR can be a viable practical solution to a particular real problem. Such a systematic and comprehensive evaluation is not possible with common clinical datasets, which have several different and unknown sources of variability that may act as confounding factors in our analysis. Therefore, for the purposes of this work, the stylized setup presented is appropriate.

A Rayleigh detection task was employed to assess the impact of the design of the signal and the depth of the DL-SR network, and a microcalcification cluster detection task was employed to study how DL-SR affects neural-network-based observers of different capacities.
The numerical results for the SKE/BKS Rayleigh detection task revealed that the loss of task-relevant information in LR images cannot be recovered by the DL-SR operation, even though mild improvement of detection performance was observed with sub-optimal observers. Furthermore, it was observed that while increasing the depth of the DL-SR network improves the traditional IQ metrics, improved task performance does not always follow.
This suggests that the mantra "deeper is better" while designing neural network architectures for image super-resolution is not necessarily applicable when task performance is considered. As such, seeking to minimize a loss function solely related to traditional image quality metrics may lead to a situation where the image statistics important to the defined task are degraded.\par
Furthermore, it is of interest to investigate conditions under which the DL-SR could improve the signal detection task performance. By using SRCNN as an example, a SKS/BKS microcalcification cluster detection task was conducted to investigate the capacity of the neural-network-based observers on super-resolved images, as compared to that on LR and HR images. It was observed that DL-SR could improve the signal detection performance of sub-optimal observers that do not accurately approximate ideal observers (IOs) due to either a limited amount of training data or the limited complexity of the observer. Given sufficient training data and an observer with sufficient complexity for the particular task considered, an IO can be approximated and the benefit of DL-SR towards improving the task performance is lost. This suggests that the impact of DL-SR on a binary signal detection task depends on a combination of factors such as the DL-SR networks, the observers and the defined task. Thus, a task-based evaluation of DL-SR methods is essential to accurately quantify the benefit of DL-SR for clinical practice. \par
Some important topics remain to be investigated in the future. The binary signal detection tasks considered in this study are simplistic compared to real-world clinical tasks. A future work could investigate the performance of DL-SR methods as preprocessing blocks on tasks such as multi-class classification, lesion segmentation and image registration. The task-based evaluation pipeline presented in this study can readily be applied to various DL-SR methods in which different network architectures or loss functions are employed. It is known that deep learning-based methods may lead to hallucinations, especially when acting on data outside the training distribution \cite{hallucination}. Hence, an objective assessment of the robustness of DL-SR methods for distribution shifts is also an important topic for future investigation. Additionally, it will be important to conduct human reader studies to assess the performance of DL-SR methods for specific clinical tasks. The results demonstrated in our study will motivate the development of DL-SR methods in directions in which the loss of task-specific information can be mitigated by incorporating such information in designing the network architecture or the loss functions \cite{task-based-ct}.

\section{Conclusion}\label{sec:conclusion}
In this study, we presented a task-based evaluation to assess the impact of deep learning-based image super-resolution methods on binary signal detection. An SKE/BKS Rayleigh detection task and an SKS/BKS microcalcification cluster detection task were conducted on simulated image datasets with a clustered lumpy background. Our results verify that the performance of an ideal observer cannot be improved via DL-SR methods, which is consistent with the data processing inequality. Also, an improvement in traditional IQ metrics induced by DL-SR does not always correlate with the impact of DL-SR on observer performance. 
Despite this, the numerical experiments presented indicate that DL-SR methods could still improve the signal detection performance of sub-optimal numerical observers in certain cases. The reported results emphasized the necessity of a task-based evaluation of DL-SR methods and suggests future avenues for developing effective DL-SR algorithms.
\subsection*{Disclosures}
The authors declare no potential conflicts of interest.
\subsection* {Acknowledgments}
This work was supported in part by NIH awards R01EB020604, R01EB023045, R01NS102213, R01CA233873, and R21CA223799. The authors greatly appreciate Michael X. Wu for proofreading the manuscript carefully and thoughtfully. Preliminary results of this work were presented at SPIE Medical Imaging 2021 and published as an SPIE Proceedings paper \cite{eval_sr_spie}. 

\bibliography{report}   
\bibliographystyle{spiejour}   




\end{spacing}
\end{document}